\documentclass[preprint,showpacs,preprintnumbers,amsmath,amssymb]{revtex4}

\usepackage{graphicx}

\usepackage{dcolumn}

\usepackage{bm}

\newcommand{\ket}[1]{|#1\rangle}
\newcommand{\bra}[1]{\langle#1|}

\begin{document}

\title{Dephasing of a flux-qubit coupled to a harmonic oscillator}

\author{P. Bertet$^1$, I. Chiorescu$^{1*}$, C.J.P.M Harmans$^1$, J.E. Mooij$^1$}

\affiliation{$^1$Quantum Transport Group, Kavli Institute of Nanoscience, Delft University of Technology, Lorentzweg $1$, $2628CJ$, Delft, The Netherlands \\
$^*$Present address : National High Magnetic Field Laboratory,
Florida State University, 1800 East Paul Dirac Drive Tallahassee,
Florida 32310, USA.}

\begin{abstract}

Decoherence in superconducting qubits is known to arise because of
a variety of environmental degrees of freedom. In this article, we
focus on the influence of thermal fluctuations in a weakly damped
circuit resonance coupled to the qubit. Because of the coupling,
the qubit frequency is shifted by an amount $n \delta \nu_0$ if
the resonator contains $n$ energy quanta. Thermal fluctuations
induce temporal variations $n(t)$ and thus dephasing. We give an
approximate formula for the qubit dephasing time as a function of
$\delta \nu_0$. We discuss the specific case of a flux-qubit
coupled to the plasma mode of its DC-SQUID detector. We first
derive a plasma mode-qubit interaction hamiltonian which, in
addition to the usual Jaynes-Cummings term, has a coupling term
quadratic in the oscillator variables coming from the
flux-dependence of the SQUID Josephson inductance. Our model
predicts that $\delta \nu_0$ cancels in certain non-trivial bias
conditions for which dephasing due to thermal fluctuations should
be suppressed.

\end{abstract}


\maketitle

\section{Introduction}

A series of recent experiments have made it clear that it is
possible to manipulate the quantum state of macroscopic electrical
circuits based on Josephson junctions
\cite{qubits,Vion02,Chiorescu03}. This breakthrough opens the way
to the realization of fundamental tests of quantum mechanics, up
to now confined to atomic physics and quantum optics, in a
solid-state physics context. An additional interest comes from the
eventual possibility of using these circuits as building blocks
for a quantum computer \cite{Makhlin02}. In view of this latest
application, it is highly desirable to understand better how they
become dephased by environmental noise.

To estimate the dephasing rates, the Bloch-Redfield theory assumes
that the qubit is weakly coupled to a bath at temperature $T$ with
a memory short compared to all relevant timescales (white noise).
In that limit, it is well known that the dephasing rate is
proportional only to the low-frequency part of the environment
spectral density. It becomes increasingly clear however that this
description is inadequate in a number of cases. For example, the
Bloch-Redfield assumptions are obviously unjustified when
dephasing is due to the fluctuations of slow environmental degrees
of freedom which typically have a $1/f$ spectrum
\cite{decoherence}. This is also the case when a resonance of
large quality factor ($Q>>1$) occurs in the environment at a
frequency comparable to the qubit, since the memory of the
environment can not be neglected then \cite{Thorwart04}. Both
processes are relevant in our experiments
\cite{Chiorescu03,Chiorescu04}. We study the quantum coherence of
a circuit called the flux-qubit, measured by a DC-SQUID. The
flux-qubit is sensitive to a number of microscopic degrees of
freedom : motion of nearby vortices trapped in superconducting
thin-films, fluctuations of the junctions critical current, and
charge noise. In addition, it is strongly coupled to the harmonic
oscillator (called HO in the remaining of this work) constituted
by the underdamped DC-SQUID and a shunt capacitor to which it is
connected to improve its resolution as a detector. In recent
experiments we observed clear signatures of the strong coupling
between the two systems, manifested by the appearance of sideband
resonances in the spectrum \cite{Chiorescu04}. In the present
article we investigate theoretically the effects of this coupling
on the qubit decoherence. Thermal fluctuations of the photon
number $n$ stored in the oscillator shift the qubit frequency by
an amount $n \delta \nu_0$ and lead to dephasing. We note that a
similar effect has been recently observed in the case of a
Cooper-pair box coupled to a waveguide resonator
\cite{Wallraff,Wallraffac}. When the resonator was driven to
perform the measurement, the qubit line was shifted and broadened
due to ac-Stark shift and photon shot noise.

In the first part of this article, we propose a simple analytical
formula giving the pure dephasing time as a function of the system
parameters. In the remaining we apply this model to the specific
case of our circuit. We start by deriving the coupling hamiltonian
between a superconducting flux-qubit and the HO. In addition to
the linear coupling term \cite{Thorwart04,Guido04}, we find it
necessary to consider the next order term which is quadratic in
the oscillator variables. We finally investigate the dependence of
the shift per photon $\delta \nu_0$ on the system bias parameters,
namely the magnetic flux enclosed by the qubit loop $\Phi_x$ and
the SQUID bias current $I_b$. In particular, we find that it is
possible to cancel the dephasing per photon $\delta \nu_0$ for
specific bias conditions, so that the influence of thermal
fluctuations on the qubit should be suppressed.

\section{Derivation of an approximate formula for the dephasing
time}

Let us consider the situation where a qubit of frequency $\nu_q$
is linearly coupled to an underdamped HO of frequency $\nu_p$ with
a coupling strength $g$. The qubit is supposed to be an ideal
undamped two-level system, whereas the HO is coupled to a bath at
temperature $T$ which damps its dynamics with a rate $\kappa$. We
are interested in the limit $\kappa<<\nu_p$ where the oscillator
is underdamped. We can write the total hamiltonian as $H=h
[-(1/2)\nu_q \sigma_z + \nu_p a^\dag a + g \sigma_x (a+a^\dag)]$,
where we introduced the Pauli matrices $\sigma_{x,y,z}$ in the
qubit Hilbert space and the usual annihilation (creation) operator
$a$ ($a^\dag$) for the HO. This is the well-studied
Jaynes-Cummings hamiltonian \cite{Jaynes_Cummings,Haroche}. Let us
recall a few results useful in the following. In the limit where
$|\delta|\equiv|\nu_q-\nu_p|>>g$ (called the dispersive regime),
the energy eigenstates of the coupled system can be written as a
function of the uncoupled energy states $\ket{i,n}$, where $i=0,1$
refers to the qubit state and $n$ to the photon state of the HO,
as \cite{Blais}

\begin{eqnarray}\label{eq:energy_states_dispersive}
  \ket{+_n}&\simeq& \ket{1,n}+ \frac{g\sqrt{n+1}}{\delta} \ket{0,n+1} \nonumber \\
  \ket{-_n}&\simeq&- \frac{g\sqrt{n+1}}{\delta} \ket{1,n}+ \ket{0,n+1}
\end{eqnarray}

their energies being

\begin{equation}\label{eq:energies_dispersive}
  E_{\pm_n}=(n+1) h \nu_p \pm \frac{h}{2} \delta \pm h
  \frac{g^2(n+1)}{\delta}
\end{equation}

One sees that because of the coupling, the energy eigenstates are
shifted by a quantity $\pm h \frac{g^2(n+1)}{\delta}$. In the
presence of $n$ photons, the dressed qubit excited state is
$\ket{+,n}$ and the ground state $\ket{-,n-1}$ so the qubit
transition is $E_{+_n}-E_{-_{n-1}}=h (\nu_q+2 g^2(n+1)/\delta)$.
This means that the qubit frequency is shifted by an amount
$\nu_{q,n}-\nu_{q,0}=n(2 g^2/\delta)=n \delta \nu_0$. Thus, any
temporal fluctuation of the photon number will lead to dephasing
\cite{Wallraffac}. Let us introduce the mean photon number in the
HO assumed to be at thermal equilibrium $\bar{n}=1/(\exp(h \nu_p/k
T)-1)$. The stationary photon number distribution is given by a
Boltzmann law $p(n)=(1/(\bar{n}+1))(\bar{n}/(\bar{n}+1))^n$. The
temporal fluctuations are characterized by the two-time
correlation function $C(\tau)=<n(0)n(\tau)>$. It is possible to
estimate $C(\tau)$ using a quantum Langevin equation approach as
in \cite{Mandel_Wolff}. We show in this way in the annex A that

\begin{equation}\label{eq:thermal_correlator}
  C(\tau)=\bar{n}(\bar{n}+1) \exp(- \kappa |\tau|) + \bar{n}^2
\end{equation}

In order to quantify the effect of these fluctuations on the qubit
coherence, we follow the analysis of Blais et al. \cite{Blais}.
The total phase accumulated by the qubit during a free evolution
is $\phi(t)=2 \pi \int_0^t \nu_q(t')dt'=\overline{\phi(t)} +
\delta \phi(t)$ where we isolated the deterministic quantity
$\overline{\phi(t)}=2 \pi (\nu_{q,0}+\bar{n}\delta \nu_0) t$ from
the fluctuating $\delta \phi(t)=2 \pi \delta \nu_0 \int_0^t
(n(t')-\bar{n})dt'$. Dephasing is described by the quantity
$f_\phi(t)=<\exp (i \delta \phi(t))>$ called the dephasing factor.
In the limit where $t>>\kappa^{-1}$, the variable $\delta \phi(t)$
should have gaussian statistics so that

\begin{equation}\label{eq:f_phi}
\begin{array}{rcl}
  f_\phi(t)&=&\exp(-<\delta \phi(t)^2>/2)\\
  &=&\exp[-(2 \pi \delta \nu_0)^2/2 \int_0^t \int_0^t
<(n(t')-\bar{n})(n(t'')-\bar{n})> dt' dt'']
\end{array}
\end{equation}

Combining equations \ref{eq:thermal_correlator} and
\ref{eq:f_phi}, we obtain that

\begin{equation}
   f_\phi(t)=\exp[-((2 \pi \delta \nu_0)^2 \bar{n}(\bar{n}+1)/2) \int_0^t \int_0^t \exp (- \kappa
   |t'-t''|)dt'dt'']
\end{equation}

Since $\int_0^t \int_0^t exp(- \kappa
|t'-t''|)=(2/\kappa)t+(2/\kappa^2)[\exp(-\kappa t)-1]$, we find
that the long-time decay of the dephasing factor is given by

\begin{equation}
  <\exp (i \delta \phi(t)>=\exp[-\frac{1}{2} (2 \pi \delta \nu_0)^2 \bar{n}(\bar{n}+1) \frac{2}{\kappa}
  t]
\end{equation}

This describes an exponential decay of time constant

\begin{equation}\label{eq:tauphi}
  \tau_\phi=\frac{\kappa}{(2 \pi \delta \nu_0)^2
  \bar{n}(\bar{n}+1)}
\end{equation}

It is interesting to compare this formula with the one derived in
\cite{Blais} for the case when the HO is driven by a coherent
field of mean photon $\bar{n}$. The photon-photon correlator is
then $C(\tau)=\bar{n} \exp{(-\kappa |t|/2)}$. Compared to equation
\ref{eq:thermal_correlator}, we notice a factor $1/2$ in the
exponent which is due to the presence of the external drive, and a
replacement of the $\bar{n}(\bar{n}+1)$ by $\bar{n}$. This
reflects the fact that a coherent field has a poissonian
distribution of photon numbers of variance $\bar{n}$ whereas a
thermal field has a superpoissonian distribution of variance
$\bar{n}(\bar{n}+1)$.

We will now show that in the limit where formula \ref{eq:tauphi}
applies (namely $\tau_\phi >> \kappa^{-1}$) the same formula also
gives the decay of the spin-echo time $\tau_E$. The reason is that
the fluctuations of the photon number occur on a much shorter
timescale than dephasing so that they can not be compensated by a
refocusing pulse. More quantitatively, the phase accumulated
during the echo sequence is

\begin{equation}
  \delta \phi_E (t) = 2 \pi \delta \nu_0 \left[ \int_0^{t/2}
  (n(t')-\bar{n})dt' -  \int_{t/2}^t (n(t')-\bar{n})dt' \right]
\end{equation}

so that the fluctuations are

\begin{eqnarray}
  <\delta \phi_E (t)^2>& = &(2 \pi \delta \nu_0)^2 [ \int_0^{t/2}\int_0^{t/2}
  (n(t')-\bar{n})(n(t'')-\bar{n})dt'dt'' \nonumber \\
  &+ & \int_{t/2}^t \int_{t/2}^t
  (n(t')-\bar{n})(n(t'')-\bar{n})dt'dt'' \nonumber \\
  & - & 2 \int_0^{t/2} \int_{t/2}^t (n(t')-\bar{n})(n(t'')-\bar{n})dt'dt''
  ]
\end{eqnarray}

Obviously we need to calculate only the last term. Using the
expression for the correlation function given earlier we find that

\begin{equation}
   \int_0^{t/2} \int_{t/2}^t
   (n(t')-\bar{n})(n(t'')-\bar{n})dt'dt''\nonumber = \frac{1}{\kappa^2} \bar{n}(\bar{n}+1) (1 - \exp(- \kappa
   t / 2))^2
\end{equation}

Combining with previous results we obtain

\begin{equation}
  <\delta \phi_E (t)^2> =(2 \pi \delta \nu_0)^2 \bar{n}(\bar{n}+1) \left[ \frac{2}{\kappa} t + \frac{4}{\kappa^2}
  (\exp(-\kappa t /2) - 1) - \frac{2}{\kappa^2} (1 - \exp(-\kappa
  t /2))^2 \right]
\end{equation}

In the limit where $t>>\kappa^{-1}$, the long-time decay is still
dominated by the term linear in $t$ and we obtain
$\tau_E=\tau_\phi$. If on the other hand we were in the opposite
limit in which the dephasing time is shorter than the photon
correlation time, the decay of the Ramsey signal would be gaussian
and the echo would decay with a $\exp(-\kappa t^3/12)$ law much
slower than the Ramsey decay. We note that this crossover between
a lorentzian and a gaussian lineshape when the dephasing time
becomes shorter than $\kappa^{-1}$ has been observed in
\cite{Wallraffac}.

Whereas in this reasoning we only considered the case where the
qubit-HO coupling is linear, it can also be applied evidently for
a more complex interaction hamiltonian whenever the qubit
frequency shift is proportional to the photon number. As we will
see in the next paragraph, this is the case for our circuit in
which a flux-qubit is coupled to the plasma mode of its measuring
DC-SQUID.

\section{Qubit-plasma mode coupling Hamiltonian}

\subsection{Description of the system}

\begin{figure}\label{fig:circuit}
\resizebox{.8\textwidth}{!}{\includegraphics{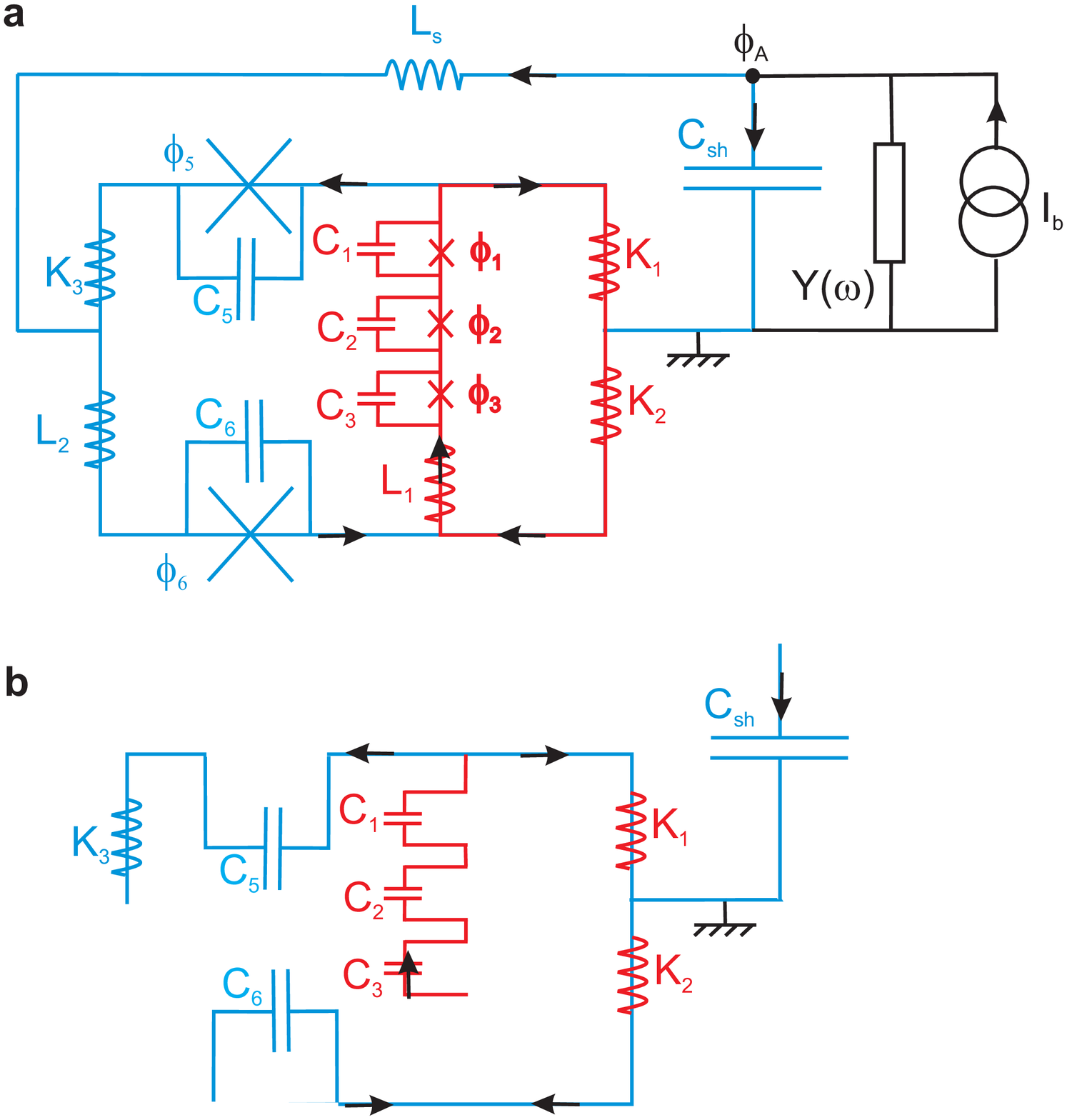}}
\caption{(a) qubit biased by $\Phi_x$ and SQUID biased by current
$I_b$. (b) Simplified electrical scheme : the SQUID-qubit system
is seen as an inductance $L_J$ connected to the shunct capacitor
$C_{sh}$ through inductance $L_{sh}$. $\Phi_a$ is the flux across
the two inductances $L_J$ and $L_{sh}$ in series.}
 \label{fig1}
\end{figure}

The flux-qubit is a superconducting loop containing three
Josephson junctions threaded by an external flux $\Phi_x \equiv f
(\Phi_0/2\pi)$ \cite{Mooij99,Caspar00,Chiorescu03}. It is coupled
to a DC-SQUID detector shunted by an external capacitor $C_{sh}$
whose role is to limit phase fluctuations across the SQUID as well
as to filter high-frequency noise from the dissipative impedance.
The SQUID is threaded by a flux $\Phi_{Sq} \equiv
f'(\Phi_0/2\pi)$. The circuit diagram is shown in figure
\ref{fig1}a. There, the flux-qubit is the loop in red containing
the three junctions of phases $\phi_i$ and capacitances $C_i$
($i=1,2,3$). It also includes an inductance $L_1$ which models the
branch inductance and eventually the inductance of a fourth larger
junction \cite{Bertet04}. The two inductances $K_1$ and $K_2$
model the kinetic inductance of the line shared by the SQUID and
the qubit. The SQUID is the larger loop in blue. The junction
phases are called $\phi_4$ and $\phi_5$ and their capacitances
$C_4$ and $C_5$. The critical current of the circuit junctions is
written $I_{Ci}$ ($i=1$ to $5$). The SQUID loop also contains two
inductances $K_3$ and $L_2$ which model its self-inductance. The
SQUID is connected to the capacitor $C_{sh}$ through
superconducting lines of parasitic inductance $L_s$. The phase
across the stray inductance and the SQUID is denoted $\phi_A$. The
whole circuit is biased by a current source $I_b$ in parallel with
a dissipative admittance $Y(\omega)$. Since our goal is primarily
to determine the qubit-plasma mode coupling hamiltonian, we will
neglect the admittance $Y(\omega)$.

We start writing the total hamiltonian of the circuit shown in
figure \ref{fig:circuit} using the circuit theory presented in
\cite{Guido03}. We first choose a spanning tree containing all the
capacitors as shown in figure \ref{fig1}a. We then write the loop
submatrices

\begin{equation}
  \mathbf{F}_{CL}=
  \left(
  \begin{array}{ccc}
  1&0&0 \\
  1&0&0 \\
  1&0&0 \\
  0&1&0\\
  0&1&-1\\
  0&0&-1
  \end{array}
  \right)
  \mbox{   , }
 \mathbf{F}_{CB}=
  \left(
  \begin{array}{c}
  0\\
  0\\
  0\\
  0\\
  0\\
  1
  \end{array}
  \right)
\end{equation}

and

\begin{equation}
\mathbf{F}_{KL}=
  \left(
  \begin{array}{ccc}
  1&-1&1\\
  1&-1&0\\
  0&1&-1
  \end{array}
  \right)
  \mbox{   , }
  \mathbf{F}_{KB}=
  \left(
  \begin{array}{c}
  0\\
  0\\
  0
  \end{array}
  \right)
  \end{equation}

We note $M$ the mutual inductance between the qubit and SQUID
loops. In the notations of \cite{Guido03} the inductance matrices
are

\begin{equation}
\mathbf{L}=
  \left(
  \begin{array}{ccc}
  L_1&M&0\\
  M&L_2&0\\
  0&0&L_3
  \end{array}
  \right)
  \mbox{   , }
  \mathbf{L}_{K}=
  \left(
  \begin{array}{ccc}
  K_1&0&0\\
  0&K_2&0\\
  0&0&K_3
  \end{array}
  \right)
  \mbox{   , }
  \mathbf{L}_{LK}=
  \left(
  \begin{array}{ccc}
  0&0&0\\
  0&0&0\\
  0&0&0
  \end{array}
  \right)
  \end{equation}

and

\begin{equation}
\mathbf{L}_{LL}=
  \left(
  \begin{array}{ccc}
  L_1+K_1+K_2&M-K_1-K_2&K_1\\
  M-K_1-K_2&L_2+K_1+K_2&-K_1-K_3\\
  K_1&-K_1-K_2&K_1+K_3+L_3
  \end{array}
  \right)
  \end{equation}

We note $l^{-1}_{ij}$ the matrix elements of $L_{LL}^{-1}$ whose
expressions can be easily computed. We finally obtain the total
hamiltonian as

\begin{equation}
   H_S=H_{kin}+(\Phi_0/2\pi)^2 U(\phi)
\end{equation}

where

\begin{equation}\label{eq:total_hamiltonian}
\begin{array}{rcl}
   H_{kin}&=&(\frac{\Phi_0}{2\pi})^2 (\frac{1}{2}\sum_{i=1}^5 Q_i^2/C_i +
   Q_A^2/C_{sh})\\
   U(\phi)&=&-\sum_{i=1}^5 \frac{1}{L_{J,i}}\cos \phi_i + \frac{1}{2l_{11}}
    (\sum_{i=1}^3 \phi_i - f)^2 \\
   & &+ (\sum_{i=1}^3 \phi_i -
   f)[l^{-1}_{21}(\phi_4+\phi_5-f')-l^{-1}_{31}(\phi_5+\phi_A)] +
   u(\phi_4,\phi_5,\phi_A)
  \end{array}
\end{equation}

The $Q_i$ ($Q_A$) are the charges stored on the capacitors $C_i$
($C_{sh}$) and $u(\phi_4,\phi_5,\phi_A)$ is defined by

\begin{equation}
\begin{array}{rcl}
   u & = & + \frac{1}{2l_{22}}  (\phi_4+\phi_5 - f')^2 - \frac{1}{2 l_{32}}
    [(\phi_5+\phi_A - f')^2) + (\phi_4+\phi_5)^2 -
   (\phi_4-\phi_A)^2]\\
   & & + \frac{1}{2 l_{33}} (\phi_5+\phi_A)^2 + \frac{2\pi}{\Phi_0} I_b
   \phi_A
  \end{array}
\end{equation}

Our first goal will be here to simplify this hamiltonian so that
the coupling of the relevant degrees of freedom is made clear. We
will consider here that they are only two : the qubit, in the
two-level approximation, and the plasma mode considered, if
uncoupled to the qubit, as a harmonic oscillator. In particular,
we will neglect the SQUID junctions capacitance which bring
additional resonances at higher frequencies, and only consider the
shunt capacitance $C_{sh}$. Our approach is justified by the fact
that only the plasma mode and the qubit have comparable energy
scales, that the plasma mode is strongly coupled to the
environment and therefore relevant for studying dephasing and
relaxation, and that it undergoes thermal fluctuations because of
its relatively low frequency. We also observed experimental
evidence for the qubit-plasma mode strong coupling
\cite{Chiorescu04}. These results are a clear indication that a
quantum-mechanical description of the coupled ``qubit-plasma mode"
is indeed needed. We will start by doing the two-level
approximation on the qubit variables.

\subsection{Qubit hamiltonian and two-level approximation}

The hamiltonian for the qubit alone is

\begin{equation}
   H_q(f,I_b)=H_{kin}-(\Phi_0/2\pi)^2[\sum_i \frac{1}{L_{J,i}}\cos \phi_i + \frac{1}{2
   l_{11}}(\sum \phi_i - f)^2]
\end{equation}

We first write this hamiltonian in a two-level approximation in
terms of the Pauli matrices, which is valid here around $\Phi_0/2$
because of the specific properties of the circuit eigenstates. We
define the states $\ket{0}$ and $\ket{1}$ as the eigenstates of
$H_q(\pi,0)\equiv H_q^{(0)}$. Then when restraining ourselves to
the $0,1$ states we have by definition $H_q^{(0)}=-(h\Delta/2)
\sigma_z$ We define $H_q^{(1)}(f,I_b)=H_q(f,I_b)-H_q^{(0)}$. We
have

\begin{equation}
\begin{array}{rcl}
   H_q^{(1)}(f,I_b)&=&(\Phi_0/2\pi)^2 [(1/2l_{11})[(\sum \phi_i - f)^2 - (\sum \phi_i -
   \pi)^2]]\\
   &=& (\Phi_0/2\pi)^2[ -(1/2l_{11}) (2\sum \phi_i - 2\pi -
   (f-\pi))(f-\pi)]\\
   &=&-(\Phi_0/2\pi)^2(\sum \phi_i - \pi)(f-\pi)/l_{11}
   \end{array}
\end{equation}

forgetting constant terms in the last equation. We now want to
decompose $H_q^{(1)}(f,I_b)$ on the $\ket{0},\ket{1}$ subspace. We
start by writing that $H_q^{(1)}(f,I_b)=(1/2)[(h_{00}+h_{11})I +
(h_{00}-h_{11})\sigma_z + (h_{01}+h_{10})\sigma_x + (i h_{01} - i
h_{10}) \sigma_y]$, where $h_{ij}=<i|H_q^{(1)}|j>$. For symmetry
reasons $<0|(\sum \phi_i - \pi)|0>=<1|(\sum \phi_i - \pi)|1>=0$ so
that $h_{00}=h_{11}=0$. Indeed, the hamiltonian $H_q^{(0)}$ is
invariant under the transformation $T$ $\phi_1\rightarrow
-\phi_1$, $\phi_2\rightarrow -\phi_2$ and $\phi_3\rightarrow 2\pi
-\phi_3$. This means that the eigenstates of $H_q^{(0)}$ have to
also be eigenstates of $T$. Since $T^2=I$, these eigenstates
should be either $1$ or $-1$ so that
$\psi_i(-\phi_1,-\phi_2,2\pi-\phi_3)=\pm
\psi_i(\phi_1,\phi_2,\phi_3)$ and
$|\psi_i(-\phi_1,-\phi_2,2\pi-\phi_3)|^2=|\psi_i(\phi_1,\phi_2,\phi_3)|^2$.
This implies that the matrix elements $h_{00}=h_{11}=0$.

Since we can always chose the global phases of $\ket{0}$ and
$\ket{1}$ so that $h_{01}$ and $h_{10}$ are real, we obtain that
$H_q^{(1)}(f,I_b)=h_{01} \sigma_x$ where
$h_{01}=(\Phi_0/2\pi)^2(\bra{0}\sum
\phi_i\ket{1})[-(f-\pi)/l_{11}]$. We next define $I_p \equiv
(\Phi_0/2 \pi) (1/l_{11}) <0|\sum \phi_i |1>$, $\epsilon=2 I_p
(\Phi_x-\Phi_0/2)/h$. In the end we can write the total
hamiltonian as

\begin{equation}\label{eq:qubit_hamiltonian}
   H_q(f)=-\frac{h}{2}(\Delta \sigma_z + \epsilon \sigma_x)
\end{equation}

\begin{figure}
\resizebox{.5\textwidth}{!}{\includegraphics{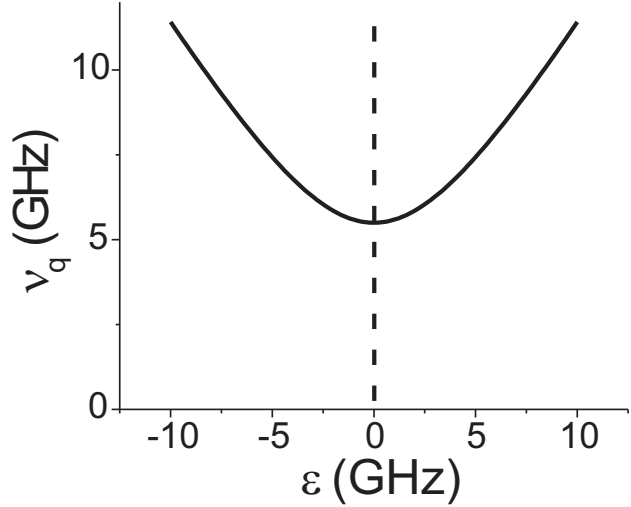}}
\caption{Qubit frequency $\nu_q$ as a function of the bias
$\epsilon$ for $\Delta=5.5GHz$ (minimum frequency in the figure).
The dashed line indicates the phase-noise insensitive bias point
$\epsilon=0$ where $d \nu_q / d \epsilon=0$ }
 \label{fig:nuq}
\end{figure}

We note that this derivation generalizes the analysis of
\cite{Orlando99} which was made under the assumption that the
qubit loop self-inductance is negligible. Here we retrieve the
result of \cite{Nakamura_theory} which showed numerically that the
form of the qubit hamiltonian was little affected by taking into
account this inductance.

The hamiltonian \ref{eq:qubit_hamiltonian} yields a qubit
transition frequency $\nu_q=\sqrt{\Delta^2 + \epsilon^2}$. The
corresponding dependence is plotted in figure \ref{fig:nuq} for
realistic parameters. An interesting property is that when the
qubit is biased at $\epsilon=0$ (dashed line in figure
\ref{fig:nuq}), it is insensitive to first order to noise in the
bias variable $\epsilon$.

\subsection{Plasma mode hamiltonian}

Next we turn to the ``SQUID+shunt capacitor" variables
$\phi_4,\phi_5,\phi_A$. As already explained, we will here make a
crude approximation and completely neglect the SQUID junctions
capacitance. This is justified by the fact that at the bias
current that we use the modes to which they correspond are at
frequencies much higher than the qubit and plasma mode. Moreover,
we will keep only the terms of second order in the SQUID
potential, which is equivalent to considering the SQUID as one
inductance $L_J(I_b,f')$. On the other hand we will keep in the
analysis the dependence on flux of $L_J$ which has important
effects.

In this approximation, the only dynamical variable in the system
is $\phi_A$. Its hamiltonian is very simply given by the
hamiltonian of a harmonic oscillator of capacitance $C_{sh}$ and
inductance $L=L_{s}+L_J(I_b,f')$

\begin{equation}
  H_{p}=Q_A^2/2C_{sh}+(\Phi_0/2\pi)^2(\phi_A-\overline{\phi_A})^2/2L
\end{equation}

where $\overline{\phi_A}=(2\pi/\Phi_0)L I_b$ is the mean value of
$\phi_A$. We call $a$ and $a^\dag$ the creation and annihilation
operators corresponding to this harmonic oscillator :

\begin{equation}
   a = \frac{\Phi_0}{2\pi}\sqrt{\frac{\pi C_{sh} \nu_0}{\hbar}}(\phi_A-\overline{\phi_A}) + \frac{i}{\sqrt{2 C_{sh} h
\nu_0}} Q_A
\end{equation}

and $a^\dag$ is the conjugate operator. Then the hamiltonian is
simply $H_p=\hbar \omega_0 (a^\dag a +1/2)$, and the phase
$\phi_A$ can be written : $\phi_A=\overline{\phi_A}+\delta \phi_0
(a+a^\dag)$ ($\delta \phi_0$ is the rms amplitude of the vacuum
fluctuations of $\phi_A$).

In our model where the SQUID junctions have no capacitance, for a
given value of $\phi_A$ and $f'$ all the phases of the SQUID are
well-defined functions $\phi_{4,5}(\phi_A,f')$. So the quantum
fluctuations of $\phi_A$ translate directly into fluctuations of
$\phi_4$ and $\phi_5$ as follows :

\begin{equation}
  \phi_{4,5}=\overline{\phi_{4,5}}+(d \phi_{4,5}/d \phi_A) \delta \phi_0
  (a+a^\dag)+(1/2)(d^2 \phi_{4,5}/d \phi_A^2) (\delta \phi_0)^2
  (a+a^\dag)^2
\end{equation}

where $\overline{\phi_{4,5}}=\phi_{4,5}(\overline{\phi_A})$. We
develop the functions to second order in $a$ and $a^\dag$ for
consistency. Again, the sensitivity coefficients $d \phi_{4,5}/d
\phi_A$ and $d^2 \phi_{4,5}/d \phi_A^2$ depend on $I_b$ and $f'$
and can be easily calculated. We also note that $d \phi_A
=(2\pi/\Phi_0) L d I_b$ so that

\begin{equation}
  \phi_{4,5}=\overline{\phi_{4,5}}+\frac{\Phi_0}{2\pi L}\frac{d \phi_{4,5}}{d I_b} \delta \phi_0 (a+a^\dag)
  +\frac{1}{2}(\frac{\Phi_0}{2 \pi L })^2\frac{d^2 \phi_{4,5}}{d I_b^2} (\delta \phi_0)^2
  (a+a^\dag)^2
\end{equation}

We will finally show how to evaluate the sensitivity coefficients
for any SQUID parameters. Following \cite{Lefevre,Balestrothesis}
we introduce the parameters $x=(\phi_4+\phi_5)/2$,
$y=(\phi_4-\phi_5)/2$, $s=I_b/(I_{C4}+I_{C5})$, $b=\Phi_0/\pi
L_{Sq} (I_{C4}+I_{C5})$, $U_0=\Phi_0 (I_{C4}+I_{C5})/2\pi$,
$\alpha=(I_{C4}-I_{C5})/(I_{C4}+I_{C5})$, and $\delta =
(K_3+K_1-L_2-K_2)/L_{Sq}$. The stationary solutions for the SQUID
phases are obtained by minimizing the $2$-dimensional potential

\begin{equation}
  U(x,y)=U_0[-sx-\cos x \cos y - \alpha \sin x \sin y -\delta s y
  + b (y-f'/2)^2]
\end{equation}

that is to solve the equations

\begin{equation}\label{eq:SQUID}
   \begin{array}{rcl}
     \partial U/\partial x&=&U_0[-s + \sin x \cos y - \alpha \cos x
     \sin y]=0 \\
      \partial U/\partial y&=&U_0[ \cos x \sin y - \alpha \sin x
     \cos y - \delta s + 2 b (y-f'/2)]=0
   \end{array}
\end{equation}

From this it is possible to obtain numerically the functions
$\Phi_{4,5}(\phi_A,f')$ and thus all the sensitivity coefficients
needed in the model.

\subsection{Qubit-plasma mode coupling hamiltonian}

The coupling hamiltonian is due to two contributions. First, the
explicit coupling term in the equation \ref{eq:hamiltonian}. But
we also need to rewrite the plasma mode hamiltonian since the
parameters of this hamiltonian (notably the SQUID Josephson
inductance) now depends on the qubit state. We therefore need to
reconsider the following variables in the plasma mode hamiltonian
:

\begin{equation}
\begin{array}{rcl}
   f'& \longrightarrow & f' + (2\pi/\Phi_0) M I_p \sigma_x \\
   L & \longrightarrow & L + \delta L \sigma_x \\
   \overline{\phi_A} & \longrightarrow & \overline{\phi_A} +
   (2\pi/\Phi_0) \delta L I_b \sigma_x
\end{array}
\end{equation}

where $\delta L \equiv (2 \pi /\Phi_0) (\partial L_J / \partial
f') M I_p$. The SQUID-qubit coupling term writes

\begin{equation}
    V=(\Phi_0/2\pi)^2(\sum_{i=1}^3 \phi_i -
   f)[l^{-1}_{21}(\phi_4+\phi_5-f')-l^{-1}_{31}(\phi_5+\phi_A)]
\end{equation}

Since $\sigma_x=(\sum \phi_i - \pi)/<0|\sum \phi_i |1>$, we can
rewrite $V$ in the form

\begin{equation}
    V=(\Phi_0/2\pi)l_{11} I_p \sigma_x
    [l^{-1}_{21}(\phi_4+\phi_5-f')-l^{-1}_{31}(\phi_5+\phi_A)]
\end{equation}

Keeping only the terms which contain explicit couplings
 we obtain that

\begin{equation}
\begin{array}{rcl}
    V&=&(\Phi_0/2\pi)l_{11} I_p \sigma_x [(\Phi_0/2 \pi L)(l^{-1}_{21}(d \phi_4/d I_b)+
     (l^{-1}_{21}-l^{-1}_{31})(d \phi_5 / d I_b) ) \delta \phi_0 (a+a^\dag)\\
    &+&(1/2)(\Phi_0/2 \pi L)^2 ( l^{-1}_{21}(d^2 \phi_4/d I_b^2)+
     (l^{-1}_{21}-l^{-1}_{31})(d^2 \phi_5 / d I_b^2)) \delta \phi_0^2 (a+a^\dag)^2]
    \end{array}
\end{equation}

On the other hand, the plasma mode hamiltonian now writes

\begin{eqnarray}
  H_{pl-q}&=&Q_A^2/2C_{sh}+(\Phi_0/2\pi)^2[\phi_A-\overline{\phi_A}-(2 \pi/\Phi_0)\delta L I_b \sigma_x]^2/2(L+\delta L \sigma_x)
  \nonumber \\
  &=&Q_A^2/2C_{sh}+\left(\frac{\Phi_0}{2\pi}\right)^2\left[\frac{(\phi_A-\overline{\phi_A})^2}{2L}-\frac{2 \pi \delta L I_b}{\Phi_0
  L}
  (\phi_A-\overline{\phi_A}) \sigma_x - \frac{\delta L }{ 2 L^2}
  (\phi_A-\overline{\phi_A})^2 \sigma_x\right]
\end{eqnarray}

so that

\begin{equation}
  H_{pl-q}=h \nu_0 (a^\dag a +1/2) - \frac{\Phi_0 \delta L I_b}{2
  \pi
  L} \delta
  \phi_0 (a+a^\dag) - \left(\frac{\Phi_0}{2\pi}\right)^2\delta \phi_0^2 (\delta L / 2 L^2)
  (a+a^\dag)^2 \sigma_x
\end{equation}

Finally the total interaction hamiltonian $H_I=V+H_{pl-q}$ can be
written as

\begin{equation}
  H_I=h[ g_1 (a+a^\dag) + g_2 (a+a^\dag)^2] \sigma_x
\end{equation}

The coupling constants $g_1$ and $g_2$ could be deduced from the
above expressions. Nevertheless we propose a way to determine them
experimentally. We first note that this coupling hamiltonian can
be rewritten

\begin{equation}
  H_I=h[ \lambda_1 \delta \phi_A + \lambda_2 \delta \phi_A^2] \sigma_x
\end{equation}

where $\delta \phi_A=\phi_A-\overline{\phi_A}$. This gives us a
very direct way of evaluating the coupling constants $g_1$ and
$g_2$ : indeed varying the bias current $I_b$ through the SQUID by
an amount $\delta I_b$ is equivalent to a variation $\delta \phi_A
= 2 \pi L \delta I_b/ \Phi_0$. Since we can experimentally measure
the dependence of the qubit frequency on the bias current
$\epsilon(I_b)$ \cite{Bertet04} and thus measure $\partial
\epsilon / \partial I_b$ and $\partial^2 \epsilon / \partial
I_b^2$, we can obtain the coupling constants {\it from the
experiment} by the following expressions :

\begin{equation}
\begin{array}{rcl}
    g_1&=&(1/2)(\partial \epsilon / \partial I_b) (\Phi_0/2 \pi L) \delta \phi_0 \\
    g_2&=&(1/4) (\partial^2 \epsilon / \partial I_b^2) (\Phi_0/2 \pi L)^2 \delta \phi_0^2
     \end{array}
\end{equation}

Finally the total qubit-plasma mode hamiltonian can be written as

\begin{equation}\label{eq:hamiltonian}
  H/h=(1/2)(-\Delta \sigma_z - \epsilon \sigma_x) +  \nu_0 (a^\dag
  a + 1/2) + [ g_1(I_b) (a+a^\dag) + g_2(I_b) (a+a^\dag)^2] \sigma_x
\end{equation}

We note that a more rigorous derivation for the coupling between
the plasma mode and the qubit after elimination of the internal
dynamics of the SQUID thanks to the Feynman-Vernon influence
functional has been performed in \cite{Nakano} and gives
ultimately the same form if we develop their interaction
hamiltonian to the second order in the oscillator variables. It is
also instructive to compare this hamiltonian to the simpler
situation studied in \cite{Wallraff}. There a Cooper-pair box is
capacitively coupled to a coplanar waveguide resonator. The
interaction hamiltonian contains only one term, linear in the
oscillator variables, and with a {\it fixed} coupling constant
depending on the geometry of the circuit. In our case the coupling
is mediated by the SQUID flux-dependent and current-dependent
inductance ; therefore the coupling constants are {\it tunable}
and higher-order terms are of importance. This made possible to
induce transitions in which both the HO and the qubit state are
modified \cite{Chiorescu04}.

\subsection{Coupling constants}

We now want to give analytical formulae for the coupling constants
$g_1$ and $g_2$ in the simplest case where a number of assumptions
are made : 1) the SQUID-qubit coupling is supposed to be symmetric
($K_1=K_2$) so that the qubit bias is only coupled to the current
$J$ circulating in the SQUID loop $\epsilon=(2 I_p/h) (\Phi_x + M
J(I_b))$ 2) the SQUID loop self-inductance and the stray
inductance $L_s$ are negligible so that the total inductance of
the plasma mode is the SQUID Josephson inductance $L=L_J(f,I_b)$.
Within these assumptions the equations \ref{eq:SQUID} are easily
solved and yield that

\begin{eqnarray}
     x&=&\arcsin\left(\frac{I_b}{2 I_C \cos (f'/2)}\right)  \nonumber \\
     y&=&f'/2
\end{eqnarray}

which implies that the current circulating in the SQUID loop is

\begin{equation}\label{eq:J}
J \equiv (I_1-I_2)/2 = I_C \sqrt{1-\left(\frac{I_b}{2 I_C \cos
(f'/2)}\right)^2} \sin (f'/2)
\end{equation}

The rms phase fluctuations of the plasma mode are $\delta \phi_0 =
(2 \pi / \Phi_0) \sqrt{h \nu_p L / 2}$ so that we obtain

\begin{eqnarray}
  g_1&=& - \frac{M I_p}{h} \frac{\sin f'/2}{4 I_C \cos^2 f'/2} \sqrt{\frac{h \nu_p}{2 L}} I_b \left[1-\left(\frac{I_b}{2 I_C \cos
  f'/2}\right)^2\right]^{-1/2}  \nonumber \\
  g_2&=& - \frac{M I_p}{h} \frac{\sin f'/2}{8 I_C \cos^2 f'/2} \frac{h \nu_p}{2 L} \nonumber \\
  & & \left[-\left(\frac{I_b}{2 I_C \cos f'/2}\right)^2 \left(1-\left(\frac{I_b}{2 I_C \cos f'/2}\right)^2\right)^{-3/2}
   + \left(1 - \left(\frac{I_b}{2 I_C \cos
   f'/2}\right)^2\right)^{-1/2}\right]
  \end{eqnarray}

Around $I_b=0$ these expressions can be simplified by keeping only
the lowest order in $I_b$ :

\begin{eqnarray} \label{eq:g1g2approx}
  g_1&\simeq& - \frac{M I_p}{h} \frac{ \sin f'/2}{4 I_C \cos^2 f'/2}  \sqrt{\frac{h \nu_p}{2 L_J}} I_b  \nonumber \\
  g_2&\simeq& - \frac{1}{16} \frac{\sin f'/2}{ \cos^2
  f'/2} \frac{M I_p}{L I_C} \nu_p
  \end{eqnarray}

\begin{figure}
\resizebox{.5\textwidth}{!}{\includegraphics{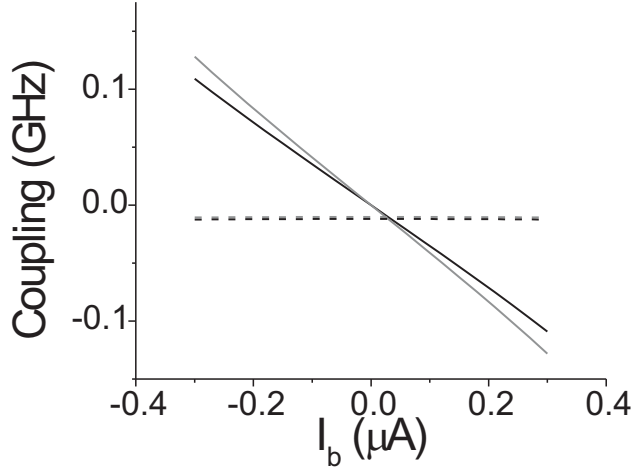}}
\caption{Coupling constants $g_1$ (solid line) and $g_2$ (dashed
line) as a function of the bias current, for two values of the
reduced SQUID flux bias $f'$ differing by $df'=4 \cdot 10^{-3}$
(gray and black lines).}
 \label{fig:couplings}
\end{figure}

We will now discuss quantitatively the behaviour of $g_1$ and
$g_2$ for actual sample parameters \cite{Bertet04} : $I_C=3.4 \mu
A$, $M=6.5pH$, $I_p=240nA$, $\Delta=5.5GHz$, $\nu_p=3.1GHz$,
$L_J=300pH$, $f'/2=1.45 \pi$. We will restrict ourselves to a
range of bias conditions relevant for our conditions, supposing
that $I_b$ varies between $\pm 300nA$ and that $f'/2$ varies by
$df'=\pm 4\cdot 10^{-3} \pi $ around $1.45 \pi$. We chose such an
interval for $f'$ because it corresponds to changing the qubit
bias point $\epsilon$ by $\pm 2 GHz$ around $0$. The constants
$g_1$ and $g_2$ are plotted in figure \ref{fig:couplings} as a
function of $I_b$ for two different values of $f'$ ($g_1$ is shown
as a full line, $g_2$ as a dashed line, and the two different
values of $f'$ are symbolized by gray for $df'=-2 \pi 4\cdot
10^{-3}$ and black for $df'=0$). It can be seen that the coupling
constants only weakly depend on the value of the flux in this
range, so that we will neglect this dependence in the following
and consider that $g_1$ and $g_2$ only depend on the bias current
$I_b$. Moreover we see from figure \ref{fig:couplings} that the
approximations made in equation \ref{eq:g1g2approx} are justified
in this range of parameters since $g_1$ is closely linear in $I_b$
and $g_2$ nearly constant. We also note that $g_1=0$ for $I_b=0$.
This fact can be generalized to the case where the SQUID-qubit
coupling is not symmetric and the junctions critical current are
dissimilar : in certain conditions these asymmetries can be
compensated for by applying a bias current $I_b^*$ for which
$g_1(I_b^*)=0$ \cite{Guido04}. At the current $I_b^*$, the qubit
is effectively {\it decoupled} from the measuring circuit
fluctuations to first order.

\section{Energy levels and dephasing}

In this paragraph we investigate the discuss the energy levels of
the hamiltonian \ref{eq:hamiltonian} and we estimate the frequency
shift per photon $\delta \nu_0$. We will only consider the case
where the qubit is detuned from the plasma mode $\nu_q-\nu_p>>g_1$
and also $|\nu_q-2\nu_p|>>g_2$.

\subsection{Energy levels}

If the qubit and the plasma mode were uncoupled (case
$g_1=g_2=0$), the system energy eigenstates would be
$\ket{i}\bigotimes\ket{n}$, where $i$ refers to the qubit state
and can be eother $0$ or $1$, and $n$ to the plasma mode
occupation number. The energy levels would simply be
$E_{i,n}^{(u)}=h(i \sqrt{\Delta^2+\epsilon^2} + n \nu_0)$. When
either $g_1$ or $g_2$ are non zero, these eigenstates are
modified, but for convenience we will still label them thanks to
the uncoupled state from which they are the closest $\ket{i,n}$.
However, the energies are now modified :
$E_{i,n}=E_{i,n}^{(u)}+\delta \nu_{i,n}$. The aim of this
paragraph is to estimate the quantity $\delta \nu_{i,n}$.

We first rewrite the hamiltonian in a more convenient way for our
purpose, by introducing the rotated axes $X$ and $Z$ defined as
$\sigma_Z=(\Delta \sigma_z + \epsilon
\sigma_x)/\sqrt{\Delta^2+\epsilon^2}$ and $\sigma_X=(-\epsilon
\sigma_z + \Delta \sigma_x)/\sqrt{\Delta^2+\epsilon^2}$. The angle
$\theta$ is defined so that $\cos \theta \equiv
\epsilon/\sqrt{\Delta^2+\epsilon^2}$ and $\sin \theta \equiv
\Delta / \sqrt{\Delta^2+\epsilon^2}$. The system hamiltonian now
writes

\begin{equation}
  H/h=-(1/2)\sqrt{\Delta^2+\epsilon^2} \sigma_Z +  \nu_p (a^\dag
  a + 1/2) + [g_1 (a+a^\dag)+g_2(a+a^\dag)^2](\cos \theta  \sigma_Z + \sin \theta
  \sigma_X)
\end{equation}

\subsubsection{Linear term}

Let us first suppose that $g_2=0$. Then the coupling is linear in
the oscillator variables, with a longitudinal component
proportional to $\cos \theta$ and a transverse component
proportional to $\sin \theta$.

We first notice that the longitudinal component has no effect on
the energy states. Indeed, the term $\sigma_Z (a+a^\dag)$ does not
shift the energy levels to first order of the perturbation theory
since $<i,n|\sigma_Z (a+a^\dag)|i,n>=0$ for all states. To second
order, all energy levels are shifted by the same quantity $(g_1
\cos \theta)^2/ \nu_p$ which implies that all the transition
frequencies stay unchanged. This conclusion stays true to all
orders of perturbation theory.

On the other hand, the transverse coupling term $g_1 \sin \theta
\sigma_X (a+a^\dag)$ produces the well-known dipersive shift in
cavity QED \cite{Raimond,Wallraff} to second-order in perturbation
theory, as calculated in the first section of this article. In the
rotating wave approximation, we recall that $\delta \nu_{i,n}^1=i
(g_1 \sin \theta)^2/\delta + (2 i -1) n (g_1 \sin
\theta)^2/\delta$ where $\delta=\sqrt{\Delta^2+\epsilon^2}-\nu_0$
is the qubit-plasma mdoe detuning. However, it is necessary here
to go beyond the rotating wave approximation since $\delta$ is of
the same order of magnitude as the qubit and the oscillator
frequency. It is easily seen that to second order of perturbation
theory, one obtains

\begin{equation}
  \delta \nu_{i,n}^1=2 i (g_1 \sin
  \theta)^2\frac{\sqrt{\Delta^2+\epsilon^2}}{\Delta^2+\epsilon^2-\nu_0^2} + (2 i -1) 2n (g_1 \sin
\theta)^2\frac{\sqrt{\Delta^2+\epsilon^2}}{\Delta^2+\epsilon^2-\nu_0^2}
\end{equation}

The first term of this equation describes the Lamb shift, which
simply renormalizes the bare qubit frequency and has no influence
on dephasing. We will thus neglect it in the following. The
frequency shift per photon is

\begin{equation}\label{eq:deltanu1}
  \delta \nu_0^1=4 (g_1 \sin
\theta)^2\frac{\sqrt{\Delta^2+\epsilon^2}}{\Delta^2+\epsilon^2-\nu_0^2}
\end{equation}

From the previous expression it is clear that the sign of $\delta
\nu_0^1$ is fully determined by the sign of $\delta$. In
particular, in our experiments where $\nu_q>\nu_p$, $\delta
\nu_0^1>0$.

\subsubsection{Quadratic term}

Next, we consider the case $g_1=0$ but $g_2>0$ (which is the case
notably at the decoupled current $I_b=I_b^*$ \cite{Guido04}). The
quadratic coupling term produces effects which are sensibly
different from the cavity QED case. Indeed, it generates a
frequency shift {\it to first order} in perturbation theory via
the term $2 g_2 \cos \theta \sigma_Z a^\dag a$. Considering that
the $g_2$ coupling is already second order, we only keep the first
order of perturbation theory. We therefore obtain that $\delta
\nu_{i,n}^2=-(2i-1) 2 g_2 n \cos \theta$ so that the shift per
photon is

\begin{equation}\label{eq:deltanu2}
  \delta \nu_0^2=- 4 g_2 \cos \theta
\end{equation}

Contrary to the shift produced by the linear coupling term, the
sign of this frequency shift now depends on $\epsilon$. Since
$g_2$ is negative (see figure \ref{fig:couplings}), $\delta
\nu_0^2$ actually has the same sign as $\epsilon$. We also note
that the quadratic term has no effect on the qubit when
$\epsilon=0$, since at that point the average flux generated by
both qubit states $\ket{0}$ and $\ket{1}$ averages out to zero so
that the SQUID Josephson inductance is unchanged.

\begin{figure}
\resizebox{.5\textwidth}{!}{\includegraphics{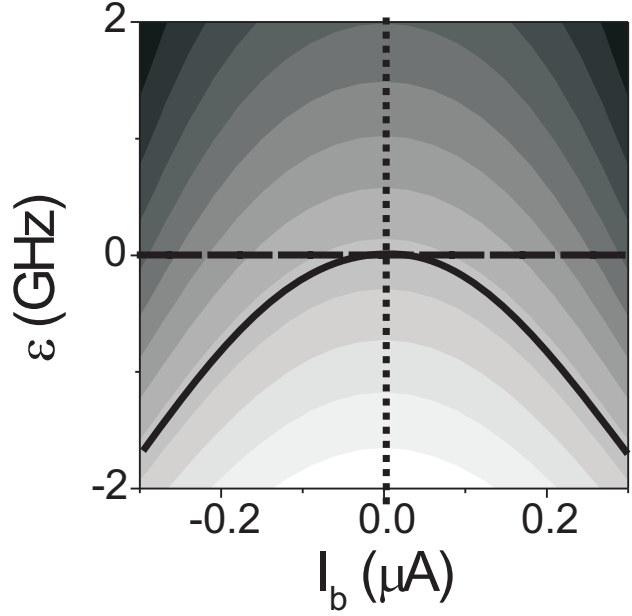}}
\caption{Frequency shift per photon $\delta \nu_0$ as a function
of $I_b$ and $\epsilon$. The white regions correspond to $-15MHz$
and the black to $+35MHz$. The solid line $\epsilon_m(I_b)$
indicates the bias conditions for which $\delta \nu_0=0$. The
dashed line indicates the phase noise insensitive point
$\epsilon=0$ ; the dotted line indicates the decoupling current
$I_b=I_b^*$.}
 \label{fig:deltanu0}
\end{figure}

\subsubsection{Total frequency shift and dependence on the bias
parameters}

The total frequency shift per photon is the sum of the two
contributions identified above :

\begin{eqnarray}\label{eq:deltanu}
  \delta \nu_0 (\epsilon,I_b) & \equiv & \delta \nu_0^1 + \delta \nu_0^2 \nonumber \\
  &=&4 \left[(g_1 \sin
\theta)^2\frac{\sqrt{\Delta^2+\epsilon^2}}{\Delta^2+\epsilon^2-\nu_p^2}-
g_2 \cos \theta \right]
\end{eqnarray}

Because of the different dependence on $\epsilon$ of the two
contributions discussed above, we expect a cancellation of the
AC-Zeeman term (due to $g_1$) by the quadratic term (due to $g_2$)
for some bias parameters corresponding to a negative value of
$\epsilon$. This is shown in figure \ref{fig:deltanu0} where we
plotted $\delta \nu_0 (\epsilon,I_b)$ as calculated with the
formula above for the sample parameters considered in the previous
paragraph. The curved full line corresponds to the points
$\epsilon_m(I_b)$ for which $\delta \nu_0=0$. For these bias
conditions, it is expected that the qubit is insensitive to the
thermal fluctuations of the plasma mode (see formula
\ref{eq:tauphi}). Therefore we predict an increase of the
dephasing time whenever $\epsilon=\epsilon_m(I_b)$.

We stress that these biasing conditions are non-trivial in the
sense that they do not satisfy an obvious symmetry in the circuit.
This point is emphasized in figure \ref{fig:deltanu0} where we
plotted as a dashed line the bias conditions $\epsilon=0$ for
which the qubit is insensitive to phase noise (due to flux or bias
current noise) ; and as a dotted line the decoupling current
conditions $I_b=I_b^*$ for which the qubit is effectively
decoupled from its measuring circuit. The $\epsilon_m(I_b)$ line
shares only one point with these two curves : the point
$(I_b^*,\epsilon)$ which is {\it optimal} with respect to flux,
bias current, and photon noise. For the rest, the three lines are
obviously distinct. This makes it possible to experimentally
discriminate between the various noise sources limiting the qubit
coherence by studying the dependence of $\tau_\phi$ on bias
parameters.

\section{Conclusion}

Superconducting qubits are often measured by circuits behaving as
underdamped oscillators to prevent energy relaxation of the qubit.
If these oscillators have a frequency comparable to $kT$, their
photon number undergoes thermal fluctuations. This induces
frequency dispersive frequency shifts of the qubit frequency $n
\delta \nu_0$ and leads to dephasing. In this article we derive a
simple fomrula to account for this process. We apply our model to
the specific case of a flux-qubit coupled to the plasma mode of
its DC-SQUID. Because of the SQUID internal degrees of freedom
(circulating current), the interaction hamiltonian contains two
terms, one linear in the oscillator variables which describes an
effective inductive coupling between the two circuit, but also a
quadratic term due to the flux-dependence of the SQUID Josephson
inductance. Moreover, the coupling constants can be tuned over a
wide range by changing the SQUID bias current. We study the qubit
frequency shift per photon $\delta \nu_0$ and find that $\delta
\nu_0=0$ for non-trivial biasing conditions. When they are
fulfilled, the effect of thermal fluctuations on the qubit should
be suppressed.

\section{Annex A}

Here we show how we evaluate the correlation function
$C(t)=<a^\dag(0) a(0) a^\dag (t) a(t)>$. In order to do so, we
follow \cite{Mandel_Wolff}. We model the damping of the HO by a
linear coupling to a bath of harmonic oscillators

\begin{equation}
\begin{array}{rcl}
  H&=&H_{HO}+H_{bath}+H_{int} \\
  &=&h \nu_p (a^\dag a +\frac{1}{2}) + \sum_\omega \hbar \omega
  (A^\dag(\omega) A(\omega)+ \frac{1}{2}) + \sum_\omega \hbar [g(\omega) a^\dag A(\omega)+g^*(\omega) A^\dag(\omega) a]
\end{array}
\end{equation}

Under the assumption that the bath has a short memory, it can be
shown that the evolution of the HO variables in the Heisenberg
representation is given by

\begin{equation}\label{eq:eqona}
  \dot{a}=(-i 2 \pi \nu_p - \kappa/2) a(t) - F(t)
\end{equation}

\noindent where $F(t)=i \sum_\omega g(\omega) A(\omega,0) \exp (-i
\omega t)$ is a quantum-mechanical operator describing the random
force acting on the HO. The damping rate of the field, which we
write $\kappa/2$ since in our notations $\kappa$ is the damping
rate of the intensity stored in the HO, can be related to the bath
parameters. Introducing the density of states $\eta(\omega)$ we
have $\kappa/2=\pi \eta ( 2 \pi \nu_p) |g( 2 \pi \nu_p)|^2$.
Integrating equation \ref{eq:eqona} we obtain

\begin{equation}\label{eq:evolution_a}
  a(t)=a(0) \exp{(-i 2 \pi \nu_p - \kappa/2)t} - \exp{(-i 2 \pi \nu_p -
  \kappa/2)t}\int_O^t F(t')\exp{(i 2 \pi \nu_p + \kappa/2)t'}dt'
\end{equation}

This allows us to calculate $C(t)=<a^\dag(0) a(0) a^\dag (t)
a(t)>$

\begin{equation}
\begin{array}{rcl}
  C(t)&=&<a^\dag(0) a(0) a^\dag (t)a(t)> \\
      &=&<n(0)^2>\exp(- \kappa t)\\
      &- &\exp{(- \kappa t)} \int_0^t <a^\dag(0) a(0) F^\dag (t') a(0)> \exp{(-i2 \pi \nu_p + (\kappa/2))t'}  \\
      &+&\exp{(- \kappa t)} <a^\dag(0) a(0) a^\dag(0) F (t')> \exp{(i 2 \pi \nu_p +
      (\kappa/2))t'}\\
     &+& \exp{(-\kappa t)} \int_0^t \int_0^t
     <a^\dag(0)a(0)F^\dag(t')F(t'')> \exp{[-i 2 \pi \nu_p (t'-t'')]}\exp
     {[(\kappa/2) (t'+t'')]}dt'dt''
   )
   \end{array}
\end{equation}

In this equation, the second and third term vanish. Indeed,

\begin{equation}
\begin{array}{rcl}
   <a^\dag(0) a(0) F^\dag (t') a(0)> & = & i \sum_\omega g(\omega)
   \exp^{- i \omega t} <a^\dag(0) a (0) a^\dag(0) A(\omega,0)> \\
  &=&i \sum_\omega g(\omega)
   \exp^{- i \omega t} <a^\dag(0) a (0) a^\dag(0)>< A(\omega,0)>
   \end{array}
\end{equation}

since it is assumed that at time $t=0$ the bath and the HO are
uncorrelated. For a bath at thermal equilibrium, $<
A(\omega,0)>=0$ so that $<a^\dag(0) a(0) F^\dag (t') a(0)>=0$. The
same reasoning holds of course to show that $<a^\dag(0) a(0)
a^\dag(0) F (t')>=0$ as well. Using the fact that $<F^\dag(t)
F(t'')>=\kappa N(2 \pi \nu_p) \delta(t'-t'')$ \cite{Mandel_Wolff},
where $N(\omega)=<A^\dag(\omega,0) A(\omega,0)>$, we can calculate
the last term :

\begin{equation}
\begin{array}{rcl}
 && \exp{(-\kappa t)} \int_0^t \int_0^t
     <a^\dag(0)a(0)F^\dag(t')F(t'')> \exp{[-i 2 \pi \nu_p (t'-t'')]}\exp
     {[(\kappa/2) (t'+t'')]}dt'dt''\\
     &=& <n(0)> \kappa N(2\pi \nu_p) \exp{(-\kappa t)} \int_0^t
     \int_0^t \delta(t'-t'') \exp{[-i 2 \pi \nu_p (t'-t'')]}
     \exp{[\kappa (t'+t'')/2]} dt' dt'' \\
     &=& <n(0)> \kappa N(2\pi \nu_p) \exp{(-\kappa t)} \int_0^t
     \exp{(\kappa t')} dt' = N(2 \pi \nu_p) <n(0)> (1 - \exp{(-\kappa
     t)})
   \end{array}
\end{equation}

Since we assume that the HO and the bath are permanently in
thermal equilibrium, $N(2 \pi \nu_p)=<n(0)>=\bar{n}$, whereas
$<n^2(0)>-<n(0)>^2=\bar{n}(\bar{n}+1)$ (non-poissonian photon
statistics of a thermal field). Therefore we obtain

\begin{eqnarray}\label{eq:C_of_t}
  C(\tau)&=& [<n^2(0)>-<n(0)>^2] \exp{(- \kappa \tau)} + <n(0)>^2 \nonumber \\
  &= & \bar{n}(\bar{n}+1) \exp{(- \kappa \tau)} + \bar{n}^2
\end{eqnarray}

\end{document}